\documentclass[a4paper, amsfonts, amssymb, amsmath, reprint, showkeys, nofootinbib, twoside, superscriptaddress, aps]{revtex4-1}
\usepackage{amsthm}
\usepackage[english]{babel}
\usepackage[utf8]{inputenc}
\usepackage[colorinlistoftodos, color=green!40, prependcaption]{todonotes}
\usepackage{mathtools}
\usepackage{physics}
\usepackage{xcolor}
\usepackage{graphicx}
\usepackage[left=23mm,right=13mm,top=35mm,columnsep=15pt]{geometry} 
\usepackage{adjustbox}
\usepackage{placeins}
\usepackage[T1]{fontenc}
\usepackage{lipsum}
\usepackage{csquotes}
\usepackage{tikz}
\usepackage{bm}
\usepackage{braket}
\usepackage{multirow}
\usepackage{qcircuit}
\usepackage{siunitx}
\usepackage{txfonts}

\usepackage{hyperref}
\hypersetup{
    colorlinks=true,
    linkcolor=blue,
    urlcolor=blue,
    citecolor=blue
    }
\usepackage[capitalise]{cleveref}

\def\Id{\mathbb{I}}

\newcommand{\R}{Ref.~}
\newcommand{\eq}{Eq.~}
\newcommand{\fig}{Fig.~}
\newcommand{\E}{\mathcal{E}}
\newcommand{\I}{\mathcal{I}}
\newcommand{\trans}{\mathrm{T}}
\newcommand{\sket}[1]{\ensuremath{\left.\left\vert #1 \right\rangle\right\rangle}}

\newcommand{\sbraket}[2]{\ensuremath{\left\langle\left\langle #1 \middle| #2 \right\rangle\right\rangle}}

\begin{document}

\title{Understanding Quantum Instruments}

\author{Akel Hashim}
    \email{ahashim@berkeley.edu}
    \affiliation{Rigetti Computing, 775 Heinz Avenue, Berkeley, CA, USA 94710}

\date{\today} 

\begin{abstract}
The quantum instrument (QI) formalism is required to model mid-circuit measurements (MCMs) and the dependence of the post-measurement state on the measurement outcome. Correctly modeling QIs is essential for applications using MCMs, such as adaptive circuits and quantum error correction. Although QIs yield a joint quantum-classical state after measurement, errors in QIs can still be represented by a $d^2 \times d^2$ superoperator (e.g., process or transfer matrix) for each outcome, just as superoperators describe Markovian errors on unitary gates. However, because the joint quantum-classical system has a distinct error model for each outcome, this complicates the usual interpretation of process- or transfer-matrix error models. This Note offers practical guidance on understanding and interpreting QI error models.
\end{abstract}

\keywords{Quantum Computing, Quantum Instruments, Error Models}

\maketitle

\section{Introduction to Quantum Instruments}\label{sec:intro}

End-of-circuit (``terminating'') measurements in quantum computations are modeled with projection-valued measures (PVMs) or, more generally, positive operator-valued measures (POVMs) \cite{PRXQuantum.6.030202}. However, the POVM formalism, which maps a quantum state to a classical probability distribution $\rho \mapsto \{ p(i|\rho) \}$, cannot describe mid-circuit measurements (MCMs), where the post-measurement state is reused in further computations. The \emph{quantum instrument} (QI) formalism addresses this by modeling quantum processes as completely positive, trace-preserving (CPTP) maps that output a joint quantum-classical state \cite{Davies1970}, $\rho \mapsto \{ [\rho_i, p(i|\rho)] \}$. QIs therefore capture \emph{measurement back action} \cite{Hatridge2013}, which is essential for predicting and understanding MCM outcomes \cite{PhysRevA.88.022107}. From a broader theoretical perspective, the QI formalism connects naturally to quantum foundations, including the modern theory of quantum measurement \cite{Ozawa1984, PhysRevA.88.022107, ji2024programmable} and quantum (in)compatibility \cite{mitra2022compatibility, mitra2023incompatibility}, as well as quantum networks \cite{Chiribella2009}.

Formally, a QI $\I$ is a CPTP map that transforms a quantum state $\rho$ as:
\begin{equation}\label{eq:QI}
    \I: \rho \mapsto \I(\rho) = \sum_i \E_i(\rho) \otimes \ketbra{i} ~,
\end{equation}
where each $\E_i(\rho)$ is a CP process which dictates the probability that a measurement outcome $i$ is observed given an input state $\rho$ (i.e., $p(i|\rho) = \Tr[\E_i(\rho)]$), and $\ketbra{i}$ is the projector associated with the classical outcome $i$. The post-measurement state (conditioned on the classical outcome $i$) is 
\begin{equation}\label{eq:post_meas_state}
    \rho_i = \frac{\E_i(\rho)}{\Tr[\E_i(\rho)]} ~.
\end{equation}
For $\I$ to be both CP \emph{and} TP, we require that the sum over all elements of the set $\{\E_i\}$ preserves total probability:
\begin{equation}\label{eq:tp}
    \Tr[\rho] = \sum_i \Tr[\E_i(\rho)] = 1 ~.
\end{equation}

While the post-measurement state $\rho_i$ is correlated with the classical outcome $i$, it is important to note that they live in different output spaces. Concretely, if the input state $\rho$ is a vector in $\mathcal{L}(\mathcal{H}_A)$, where $\mathcal{H}_A = \mathbb{C}^d$ and $d=D^n$ (for $n$ qudits of dimension $D$) is the corresponding Hilbert space, then the quantum instrument transforms $\mathcal{L}(\mathcal{H}_A)$ as:
\begin{equation}
    \I: \mathcal{L}(\mathcal{H}_A) \mapsto \mathcal{L}(\mathcal{H}_B) \otimes \mathcal{L}(\mathcal{H}_K) ~,
\end{equation}
where $\mathcal{H}_K = \mathbb{C}^{\abs{K}}$ is the Hilbert space of the classical register with $\abs{K}$ possible measurement outcomes. For example, for performing a projective measurement on a single-qubit input state, $\mathcal{H}_A = \mathcal{H}_B = \mathcal{H}_K = \mathbb{C}^2$. Note that, even though the classical outcomes belong to $\mathbb{Z}_d$, we cannot represent the Hilbert space of the classical register by $\mathbb{Z}_d$, since it is not a complex inner product space (e.g., we cannot multiply elements of $\mathbb{Z}_d$ by $i$ and remain in $\mathbb{Z}_d$)\footnote{However, $\ell^2(Z_d)$ --- the space of square-integrable complex-valued functions on $\mathbb{Z}_d$ --- is a Hilbert space.}. So, even though the classical outcomes themselves are not complex, in order for $\I$ to be a valid CPTP map, we require that $\mathcal{H}_B \otimes \mathcal{H}_K$ be a valid Hilbert space, thus $\mathcal{H}_K = \mathbb{C}^{|K|} = \mathbb{C}^d$. What distinguishes $\mathcal{H}_K$ from $\mathcal{H}_B$ is that states in $\mathcal{H}_K$ must be diagonal in the chosen basis (this is captured by the projector $\ketbra{i}$ in \eq\ref{eq:QI}); i.e., no coherences are preserved.

Finally, suppose that we have back-to-back MCMs, where the first MCM is represented by the QI
\begin{equation}
    \I_1(\rho) = \sum_i \E_i(\rho) \otimes \ketbra{i}_{K_1}
\end{equation}
and the second MCM is represented by the QI
\begin{equation}
    \I_2(\sigma) = \sum_j \mathcal{F}_j(\sigma) \otimes \ketbra{j}_{K_2} ~.
\end{equation}
When we apply $\I_2$ to $\I_1(\rho)$, $\I_2$ only acts on the quantum part of $\I_1(\rho)$ --- the classical register in $H_{K_1}$ remains untouched, but must be accounted for in the final outcomes. Concretely,
\begin{align}
    \I_2: \I_1(\rho) &\mapsto (\I_2 \otimes I_{K_1}) \circ \I_1(\rho) ~, \\
    &= \sum_i I_2\left( \E_i(\rho) \right) \otimes \ketbra{i}_{K_1} ~, \\
    &= \sum_{i,j} (\mathcal{F}_j \circ \E_i)(\rho) \otimes \ketbra{j}_{K_2} \otimes \ketbra{i}_{K_1} ~.
\end{align}
Now, trace-preservation is given by $\sum_{i,j} (\mathcal{F}_j \circ \E_i)(\rho) = 1$, and the classical register resides in the combined Hilbert space $\mathcal{H}_{K_1} \otimes \mathcal{H}_{K_2} = \mathbb{C}^{|K_1|\cdot|K_1|}$. As we can see, the number of possible classical outcomes (and thus the number of possible post-measurement states) grows exponentially with the number of MCMs.
\section{Modeling Errors with Pauli Transfer Matrices}\label{sec:ptm}

\begin{figure}[t]
    \centering
    \includegraphics[width=\columnwidth]{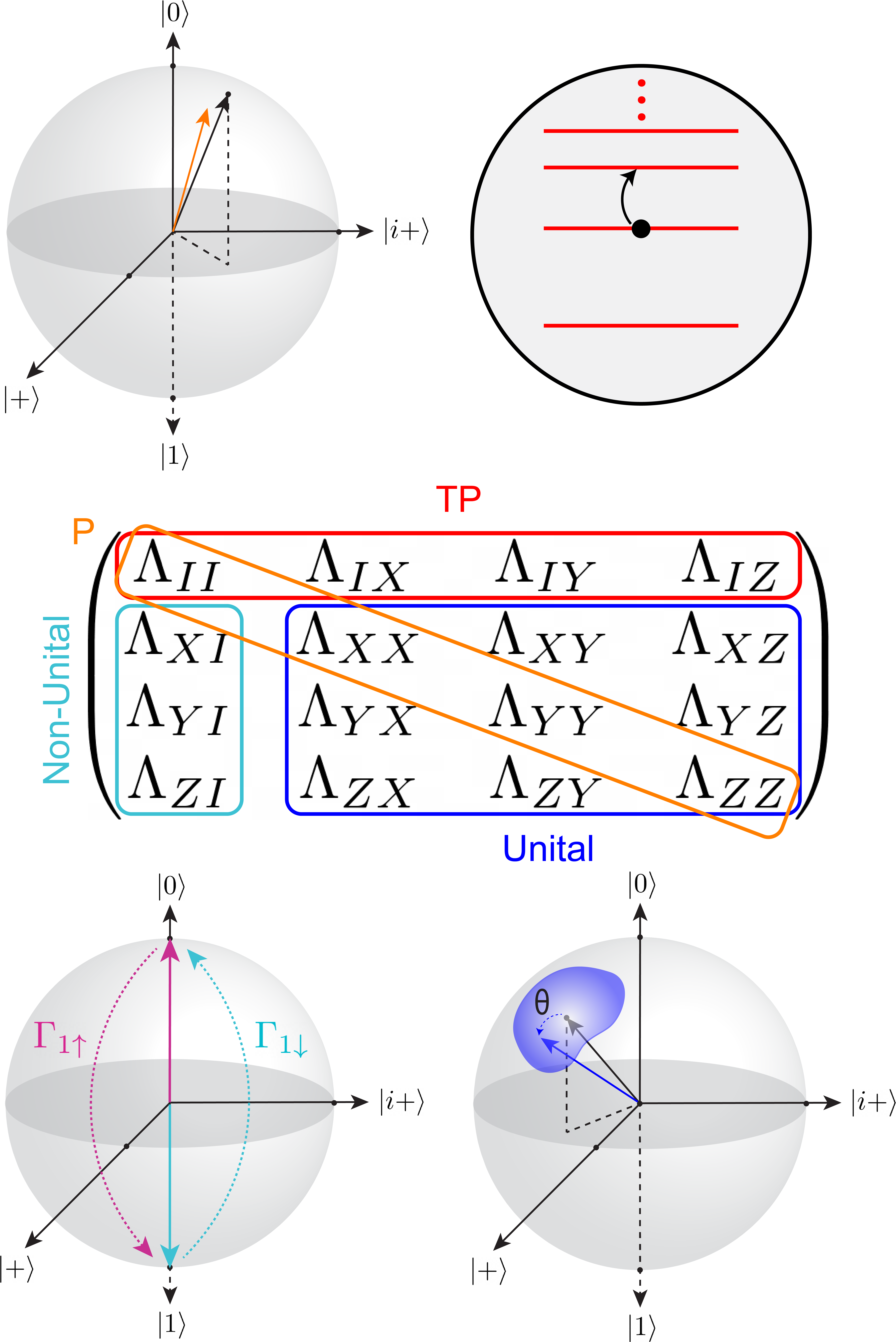}
    \caption[Pauli Transfer Matrix.]{\textbf{Pauli Transfer Matrix.} 
    We can identify four useful blocks within a PTM. The top row is typically fixed as $[1, 0, 0, 0]$ by trace preservation (TP, red), although postselected operations can be non-TP. The lower right-hand block (blue) captures unital processes, such as unitary errors. The column to the left of the unital block (cyan) indicates non-unital processes, such as $T_1$ decay, resulting in $\Lambda_{PI} \ne 0$ for $P \in \{X, Y, Z\}$. The diagonal elements indicate how well polarization (P, orange) along the various Pauli axes is preserved, and are directly impacted by stochastic Pauli noise. The spheres at each corner depict the impact that example errors captured by each block have on a Bloch vector in the Bloch sphere. (Figure and caption reprinted with permission from \R\cite{PRXQuantum.6.030202}.)
    }
    \label{fig:ptm}
\end{figure}

Because QIs can be represented as CPTP processes, errors in QIs can be modeled in the same manner as errors in unitary operations, that is, in terms of linear superoperators\footnote{This section is (largely) reprinted with permission from \R\cite{PRXQuantum.6.030202}.}.
Common superoperator representations include the transfer-matrix representation, the $\chi$-matrix representation (commonly referred to as ``process matrices''), and the Choi-matrix representation \cite{PRXQuantum.6.030202, greenbaum2015introduction, Wood2015}. 
Here, we focus on the transfer-matrix representation of errors (denoted $\Lambda$), which acts on vectorized input states $\ket{\rho\rangle}$ by direct matrix multiplication,
\begin{equation}
    \E: \ket{\rho\rangle} \mapsto \ket{\E(\rho)\rangle} = \Lambda \ket{\rho\rangle} ~.
\end{equation}
This representation is convenient because the composition of multiple quantum operations is associative. 
That is, if two processes ($\Lambda_1$ and $\Lambda_2$) are applied in succession, then the total process is just the product of the two matrices:
\begin{equation}
    \E_2, \E_1: \ket{\rho\rangle} \mapsto \ket{\E_2(\E_1(\rho))\rangle} = \Lambda_2 \ket{\E_1(\rho)\rangle} = \Lambda_2 \Lambda_1 \ket{\rho\rangle} ~.
\end{equation}

When we construct $\Lambda$ in the Pauli basis, we call this the \emph{Pauli transfer matrix} (PTM). 
This is convenient for many reasons, in particular because many error processes are commonly represented in the Pauli basis (e.g., stochastic bit- or phase-flips) throughout the literature. 
The PTM $\Lambda$ of an error process $\E$ is a $d^2 \times d^2 = 4^n \times 4^n$ superoperator with entries
\begin{equation}
    \Lambda_{ij} = \frac{1}{d}\Tr \left[ P_i\E(P_j) \right] ~,
\end{equation}
where $P_i$ and $P_j$ are elements of the $n$-qubit Pauli group $\mathbb{P}_n = \{I, X, Y, Z\}^{\otimes n}$. 
Because it is a transfer matrix, a PTM acts on density matrices by direct matrix multiplication, 
\begin{equation}\label{eq:ptm_matrix_mult}
    \ket{\rho'\rangle} = \ket{\E(\rho)\rangle} = \Lambda \sket{\rho} ~,
\end{equation}
where
\begin{equation}
    \sket{\rho} = \begin{pmatrix} \rho_{I^{\otimes n}} & ... & \rho_{Z^{\otimes n}} \end{pmatrix}^\trans
\end{equation}
is the vectorized density matrix consisting of the expansion coefficients $\rho_P = \sbraket{P}{\rho} / d$ of $\rho$ expanded in the Pauli basis,
\begin{equation}
    \rho = \sum_{P \in  \mathbb{P}_n} \rho_P P ~.
\end{equation}
For example, for a single qubit state $\rho$, the vectorized representation is
\begin{equation}
    \sket{\rho} = \begin{pmatrix}
                        \rho_{I} \\
                        \rho_{X} \\
                        \rho_{Y} \\
                        \rho_{Z}
                   \end{pmatrix} ~,
\end{equation}
and \eq\eqref{eq:ptm_matrix_mult} can be written explicitly as
\begin{equation}
    \begin{pmatrix}
        \rho_{I}' \\
        \rho_{X}' \\
        \rho_{Y}' \\
        \rho_{Z}
    \end{pmatrix} = 
    \begin{pmatrix}
        \Lambda_{II} & \Lambda_{IX} & \Lambda_{IY} & \Lambda_{IZ} \\
        \Lambda_{XI} & \Lambda_{XX} & \Lambda_{XY} & \Lambda_{XZ} \\
        \Lambda_{YI} & \Lambda_{YX} & \Lambda_{YY} & \Lambda_{YZ} \\
        \Lambda_{ZI} & \Lambda_{ZX} & \Lambda_{ZY} & \Lambda_{ZZ}
    \end{pmatrix}
    \begin{pmatrix}
        \rho_{I} \\
        \rho_{X} \\
        \rho_{Y} \\
        \rho_{Z}
    \end{pmatrix} ~.
\end{equation}

The PTM representation is useful because all of its entries are real and bounded by $\Lambda_{ij} \in [-1, 1]$, and different aspects of a process can be discerned directly by visual inspection of the matrix. 
We highlight four (partially overlapping) blocks within a PTM, as shown in \fig\ref{fig:ptm}:
\begin{itemize}
    \item $\Lambda$'s top row reveals whether it is trace-preserving. 
    The operation is TP if and only if $\Lambda_{0j} = \delta_{0j}$ (i.e., if the first row of the PTM is $(1, 0, ..., 0)$). 
    Every deterministic process must be TP, but postselected operations provide an example of non-TP processes\footnote{
    Note that some authors consider leakage, for example, to be a non-TP process, in which case the top row of the PTM captures state-dependent leakage. 
    This is true if one only considers the qubit subspace within the full Hilbert space. 
    However, strictly speaking, leakage is still TP, since the total probability of observing \emph{some} outcome is preserved. 
    For example, in some platforms leakage cannot be detected, and might instead be (erroneously) measured as 0 or 1, but the total number of shots will remain the same. 
    In other platforms leakage can be more easily measured, in which case the total probability of observing 0, 1, or 2 is preserved. 
    Therefore, when considering only the qubit subspace in the presence of leakage, it is sometimes common to relax the TP constraint, and instead simply require that the total probability must not increase (i.e., $\Tr[\E(\rho)] \le \Tr[\rho]$).}.
    
    \item The bottom right $(d^2-1) \times (d^2-1)$ block of the PTM is called the \emph{unital} block. 
    An operation is unital if it preserves the identity [i.e., $\E(\Id) = \Id$], and the PTM for a unital operation is restricted to this block and the $1\times 1$ block defined by $\Lambda_{00}$. 
    Unital processes cannot increase purity (or decrease entropy). 
    Unitary dynamics (e.g., coherent errors) and stochastic Pauli errors are unital.
    
    \item The leftmost column of $\Lambda$ is called the \emph{non-unital} block. 
    For any unital operation, $\Lambda_{i0} = \delta_{i0}$ (i.e., the first column of the PTM is $(1, 0, ..., 0)^\trans$). 
    If it does not take this form, its elements indicate entropy-decreasing processes like cooling, energy relaxation, or spontaneous emission (e.g., $T_1$ decay). 
    
    \item The diagonal elements of $\Lambda$ quantify how well \emph{polarization} is preserved along each Pauli axis (i.e., how well a quantum state prepared in $P$ and measured in $P$ is preserved), with $\Lambda_{PP} = 1$ if the operation preserves the component of Pauli operator $P$ in $\rho$. 
    $\Lambda_{PP} < 1$ indicates loss of polarization or coherence along a Pauli axis.
    A PTM's diagonal elements $\Lambda_{PP}$ are sometimes called the \emph{survival probabilities}, \emph{Pauli fidelities}, or (when $\Lambda$ is diagonal) \emph{Pauli eigenvalues}. An operation's PTM is diagonal if and only if it is a Pauli channel.
\end{itemize}
It is simple to enforce trace-preservation in the PTM representation, by requiring that $\Lambda_{0j} = \delta_{0j}$. In contrast, the CP constraint is hard to express or evaluate in the PTM representation. The easiest way to test whether a PTM $\Lambda$ describes a CP map is to construct its Choi matrix representation. In the following section, we discuss how to interpret errors in quantum instruments using PTMs.

\section{Errors in Quantum Instruments}\label{sec:qi}

\begin{figure*}[t]
    \centering
    \includegraphics[width=1.5\columnwidth]{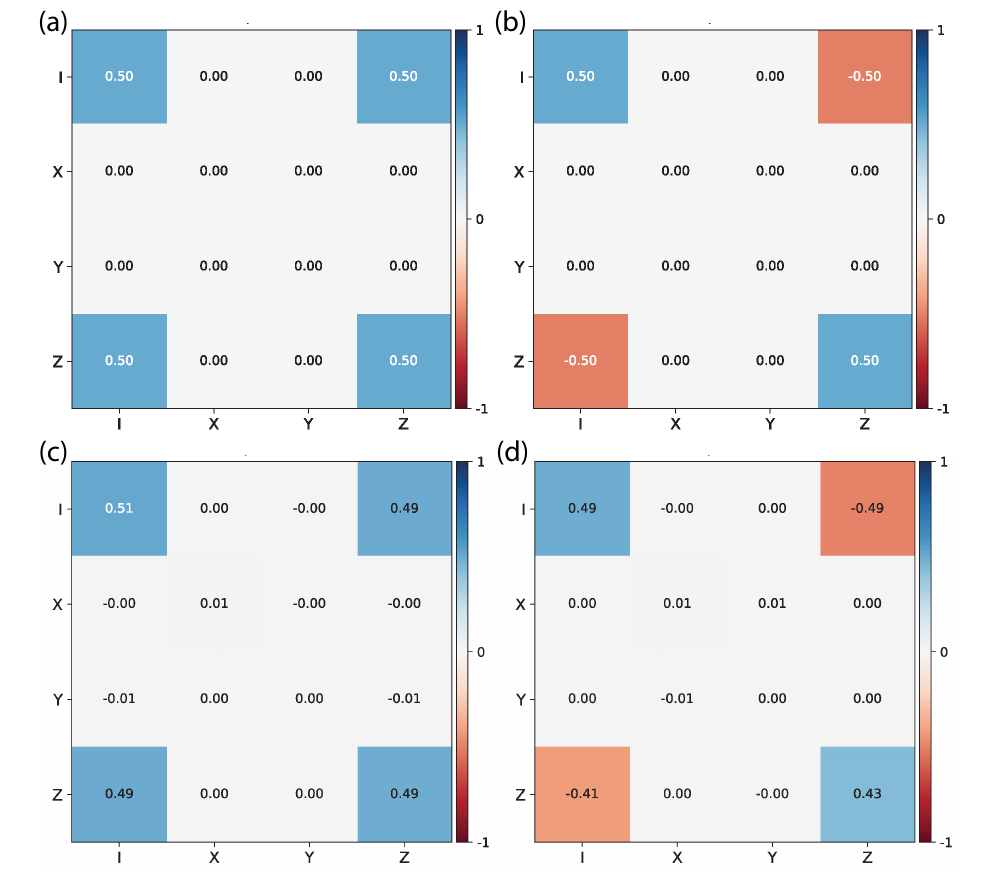}
    \caption[Quantum Instrument PTMs.]{\textbf{Quantum Instrument PTMs.}
    \textbf{(a)} Ideal (target) PTM for the measure 0 element of a single-qubit QI.
    \textbf{(b)} Ideal (target) PTM for the measure 1 element of a single-qubit QI.
    \textbf{(c)} Experimental PTM for the measure 0 element of a single-qubit QI.
    \textbf{(d)} Experimental PTM for the measure 1 element of a single-qubit QI.
    }
    \label{fig:qi}
\end{figure*}

Quantum instruments represent valid CPTP processes, but they decompose into $d$ different CP processes whose sum must be TP. This complicates the direct interpretation of the PTM representation of errors introduced in the previous section. Moreover, QIs do not just model quantum errors, but must also capture purely classical effects such as \emph{readout infidelity} (i.e., \emph{assignment error}), which can be due to quantum effects (e.g., $T_1$ decay) or classical effects (e.g., shot noise, poor signal-to-noise ratio, etc.). 
Since a QI decomposes into a joint quantum-classical state after measurement, 
it requires a modification in how we typically interpret errors in the PTM representation of CPTP processes.

To begin with, we take the simple example of a QI for a single-qubit system (i.e., a 2-outcome MCM):
\begin{equation}
    \I(\rho) = \E_0(\rho) \otimes \ketbra{0} + \E_1(\rho) \otimes \ketbra{1} ~.
\end{equation}
We denote the PTMs of $\E_0$ and $\E_1$ as $\Lambda^{\scriptscriptstyle(0)}$ and $\Lambda^{\scriptscriptstyle(1)}$, respectively. Using the notation introduced in the previous section, we may write them out as
\begin{equation}
    \renewcommand{\arraystretch}{1.5}
    \Lambda^{\scriptscriptstyle(0)} = \begin{pmatrix}
        \Lambda^{\scriptscriptstyle(0)}_{II} & \Lambda^{\scriptscriptstyle(0)}_{IX} & \Lambda^{\scriptscriptstyle(0)}_{IY} & \Lambda^{\scriptscriptstyle(0)}_{IZ} \\
        \Lambda^{\scriptscriptstyle(0)}_{XI} & \Lambda^{\scriptscriptstyle(0)}_{XX} & \Lambda^{\scriptscriptstyle(0)}_{XY} & \Lambda^{\scriptscriptstyle(0)}_{XZ} \\
        \Lambda^{\scriptscriptstyle(0)}_{YI} & \Lambda^{\scriptscriptstyle(0)}_{YX} & \Lambda^{\scriptscriptstyle(0)}_{YY} & \Lambda^{\scriptscriptstyle(0)}_{YZ} \\
        \Lambda^{\scriptscriptstyle(0)}_{ZI} & \Lambda^{\scriptscriptstyle(0)}_{ZX} & \Lambda^{\scriptscriptstyle(0)}_{ZY} & \Lambda^{\scriptscriptstyle(0)}_{ZZ}
    \end{pmatrix}
\end{equation}
and 
\begin{equation}
    \renewcommand{\arraystretch}{1.5}
    \Lambda^{\scriptscriptstyle(1)} = \begin{pmatrix}
        \Lambda^{\scriptscriptstyle(1)}_{II} & \Lambda^{\scriptscriptstyle(1)}_{IX} & \Lambda^{\scriptscriptstyle(1)}_{IY} & \Lambda^{\scriptscriptstyle(1)}_{IZ} \\
        \Lambda^{\scriptscriptstyle(1)}_{XI} & \Lambda^{\scriptscriptstyle(1)}_{XX} & \Lambda^{\scriptscriptstyle(1)}_{XY} & \Lambda^{\scriptscriptstyle(1)}_{XZ} \\
        \Lambda^{\scriptscriptstyle(1)}_{YI} & \Lambda^{\scriptscriptstyle(1)}_{YX} & \Lambda^{\scriptscriptstyle(1)}_{YY} & \Lambda^{\scriptscriptstyle(1)}_{YZ} \\
        \Lambda^{\scriptscriptstyle(1)}_{ZI} & \Lambda^{\scriptscriptstyle(1)}_{ZX} & \Lambda^{\scriptscriptstyle(1)}_{ZY} & \Lambda^{\scriptscriptstyle(1)}_{ZZ}
    \end{pmatrix} ~.
\end{equation}
$\Lambda^{\scriptscriptstyle(0)}$ ($\Lambda^{\scriptscriptstyle(1)}$) models the quantum process that is conditional on observing the measurement outcome 0 (1), and $\Lambda^{\scriptscriptstyle(0)} + \Lambda^{\scriptscriptstyle(1)}$ models the quantum process in which the classical outcome is discarded after measurement.

\subsection{Trace Preservation}

Because each $\E_i$ is not TP, there is no requirement that $\Lambda^{\scriptscriptstyle(0)}_{0j} = \Lambda^{\scriptscriptstyle(1)}_{0j} = \delta_{0j}$\footnote{In the following subsections, we use $\Lambda^{\scriptscriptstyle(i)}_{0j}$ to denote the first row of $\Lambda^{\scriptscriptstyle(i)}$ when summation notation is convenient, and $\Lambda^{\scriptscriptstyle(i)}_{I}$ otherwise.}. 
Instead, we only require that $\Lambda^{\scriptscriptstyle(0)}_{0j} + \Lambda^{\scriptscriptstyle(1)}_{0j} = \delta_{0j}$. 
This TP constraint means that $\Lambda^{\scriptscriptstyle(0)}_{I}$ and $\Lambda^{\scriptscriptstyle(1)}_{I}$ are not linearly independent (more on this below). 
In \fig\ref{fig:qi}, we plot the PTMs $\Lambda^{\scriptscriptstyle(0)}$ and $\Lambda^{\scriptscriptstyle(1)}$ for an ideal QI, as well as the experimental PTMs measured via QI linear gate-set tomography (QILGST) \cite{rudinger2022characterizing}. 
We observe that in the ideal case, $\Lambda^{\scriptscriptstyle(0)}_{I} = (0.5, 0, 0, 0.5)$ and $\Lambda^{\scriptscriptstyle(1)}_{I} = (0.5, 0, 0, -0.5)$, but experimentally we measure $\Lambda^{\scriptscriptstyle(0)}_{I} = (0.51, 0, 0, 0.49)$ and $\Lambda^{\scriptscriptstyle(1)}_{I} = (0.49, 0, 0, -0.49)$.
Thus, although there are errors in our QI, this process is indeed TP since $\Lambda^{\scriptscriptstyle(0)}_{I} + \Lambda^{\scriptscriptstyle(1)}_{I} = (0.51, 0, 0, 0.49) + (0.49, 0, 0, -0.49) = (1, 0, 0, 0)$.

While PTMs model \emph{quantum} processes, the classical output of a QI is encoded in the top row of each PTM.
To see this, we may calculate the probability of outcome $i$ given an input state $\sket{\rho}$ as
\begin{equation}\label{eq:born_QI}
    p(i | \rho) = \Lambda^{\scriptscriptstyle(i)}_{I} \sket{\rho} = \sum_j \Lambda^{\scriptscriptstyle(i)}_{0j} r_j ~,
\end{equation}
where $r_j = (1, r_x, r_y, r_z)$.
For example, if a qubit is prepared in the ground state, $\sket{\rho} = \sket{\ketbra{0}} = (1, 0, 0, 1)^\trans$, the probabilities of measuring 0 and 1 are (according to the target model)
\begin{align}
    p(0|0) &= \Lambda^{\scriptscriptstyle(0)}_{I}\sket{\ketbra{0}} = \Lambda^{\scriptscriptstyle(0)}_{II} + \Lambda^{\scriptscriptstyle(0)}_{IZ} = 0.5 + 0.5 = 1 ~, \\
    p(1|0) &= \Lambda^{\scriptscriptstyle(1)}_{I}\sket{\ketbra{0}} = \Lambda^{\scriptscriptstyle(1)}_{II} + \Lambda^{\scriptscriptstyle(1)}_{IZ} = 0.5 - 0.5 = 0 ~.
\end{align}
Experimentally, we indeed observe that $p(0|0) = 1$ and $p(1|0) = 0$ (up to any rounding error in the values displayed in the PTMs).
On the other hand, if a qubit is prepared in the excited state, $\sket{\rho} = \sket{\ketbra{1}} = (1, 0, 0, -1)^\trans$, the probabilities of measuring 0 and 1 are (according to the target model)
\begin{align}
    p(0|1) &= \Lambda^{\scriptscriptstyle(0)}_{I}\sket{\ketbra{1}} = \Lambda^{\scriptscriptstyle(0)}_{II} - \Lambda^{\scriptscriptstyle(0)}_{IZ} = 0.5 - 0.5 = 0 ~, \\
    p(1|1) &= \Lambda^{\scriptscriptstyle(1)}_{I}\sket{\ketbra{1}} = \Lambda^{\scriptscriptstyle(1)}_{II} - \Lambda^{\scriptscriptstyle(1)}_{IZ} = 0.5 + 0.5 = 1 ~.
\end{align}
Experimentally, we instead observe that $p(0|1) = 0.02$ and $p(1|1) = 0.98$ (up to any rounding error in the values displayed in the PTMs). Thus, while we observe errors in the conditional outcome probabilities when a qubit is prepared in the excited state (likely due to errors such as $T_1$ decay), total probability is still preserved: $p(0|1) + p(1|1) = 1$.

This analysis shows that the TP constraint across all elements of a QI enforces a linear dependence between $\Lambda^{\scriptscriptstyle(0)}_{I}$ and $\Lambda^{\scriptscriptstyle(1)}_{I}$. Namely, if
\begin{equation}
    \Lambda^{\scriptscriptstyle(0)}_{I} = \left( \Lambda^{\scriptscriptstyle(0)}_{II}, \Lambda^{\scriptscriptstyle(0)}_{IX}, \Lambda^{\scriptscriptstyle(0)}_{IY}, \Lambda^{\scriptscriptstyle(0)}_{IZ} \right) ~,
\end{equation}
then 
\begin{equation}
    \Lambda^{\scriptscriptstyle(1)}_{I} = \left( 1 - \Lambda^{\scriptscriptstyle(0)}_{II}, -\Lambda^{\scriptscriptstyle(0)}_{IX}, -\Lambda^{\scriptscriptstyle(0)}_{IY}, -\Lambda^{\scriptscriptstyle(0)}_{IZ} \right) ~,
\end{equation}
such that $\Lambda^{\scriptscriptstyle(0)}_{I} + \Lambda^{\scriptscriptstyle(1)}_{I} = (1, 0, 0, 0)$.

\subsection{Classical Readout Fidelity}

When performing measurements (either terminating or mid-circuit), the first question that many ask is, ``what is my readout fidelity?''
Classical readout fidelity is typically quantified using an assignment fidelity (or \emph{confusion}) matrix $\mathcal{M}$ constructed from preparing a qubit in $\ket{0}$ and $\ket{1}$ and measuring the resulting probabilities of obtaining the classical outcomes 0 and 1 in both cases:
\begin{equation}
    \mathcal{M} = \begin{pmatrix}
        p(0|0) & p(0|1) \\
        p(1|0) & p(1|1)
    \end{pmatrix} ~.
\end{equation}
$\mathcal{M}$ is a classical stochastic matrix that only quantifies measurement errors for computational basis input states and measurements in the computational basis\footnote{It is often assumed that confusion matrices quantify computational basis \emph{measurement} errors (and this is generally a good assumption), but unless an experimenter can prepare perfect initial states (we cannot), it is not possible to unambiguously separate errors in state preparation from those in measurement. For this reason, it actually captures both state-preparation and measurement (SPAM) errors in the computational basis.}.
However, they cannot generally capture measurement errors on quantum superposition states, unless errors during readout are purely incoherent \cite{beale2023randomized, hashim2025quasi}.

Analyzing the TP constraint of experimental PTMs of QIs shows that we can directly extract a confusion matrix from the measured probabilities for the input states $\rho = \ketbra{0}$ and $\rho = \ketbra{1}$:
\begin{equation}
    \renewcommand{\arraystretch}{1.5}
    \mathcal{M} =
    \begin{pmatrix}
        \Lambda^{\scriptscriptstyle(0)}_{II} + \Lambda^{\scriptscriptstyle(0)}_{IZ} & \Lambda^{\scriptscriptstyle(0)}_{II} - \Lambda^{\scriptscriptstyle(0)}_{IZ} \\
        \Lambda^{\scriptscriptstyle(1)}_{II} + \Lambda^{\scriptscriptstyle(1)}_{IZ} & \Lambda^{\scriptscriptstyle(1)}_{II} - \Lambda^{\scriptscriptstyle(1)}_{IZ}
    \end{pmatrix} ~.
\end{equation}
This is useful for comparing standard readout fidelities modeled by confusion matrices to the actual readout fidelities in a MCM. 
For example, for our experimental data, 
\begin{equation}
    \mathcal{M} = \begin{pmatrix}
        p(0|0) & p(0|1) \\
        p(1|0) & p(1|1)
    \end{pmatrix} 
    \approx \begin{pmatrix}
        1 & 0.02 \\
        0 & 0.98
    \end{pmatrix} ~.
\end{equation}
This corresponds to a classical assignment (readout) fidelity of $\mathcal{F} = [p(0|0) + p(1|1)] / 2 \approx 0.99$.
However, QIs go further than a simple confusion matrix model of readout errors, since they can capture readout errors in superposition states without making any assumptions about the details of the error model during readout (except that errors are Markovian). 
But this comes at the expense of a larger experimental overhead (an $n$-qubit confusion matrix only requires preparing $d = 2^n$ input states, whereas measurement tomography requires preparing $d^2$ input states)\footnote{It should be noted that QILGST is relatively lightweight, only requiring 128 circuits to fully reconstruct a QI \cite{rudinger2022characterizing}. 
This is more expensive than measuring a 2-qubit confusion matrix, but still very experimentally tractable, while providing much more information/insight into the nature of errors in MCMs.}.
Nevertheless, because QIs are modeled as CPTP channels, one can define suitable error metrics for quantum processes such as entanglement (process) infidelity and diamond distance for QIs \cite{mclaren2022evaluation, mclaren2023stochastic}, which capture much more about the behavior of MCMs than readout fidelity alone.

\subsection{Measurement Axis}

An ideal computational-basis measurement is along the $Z$-axis.
However, it is experimentally possible to measure along another axis \cite{HacohenGourgy2016}, either by choice or by accident.
The measurement axis of a QI is encoded in the first row of the component PTMs.
For example, for measurements along the $X$-axis the nonzero elements of each $\Lambda^{\scriptscriptstyle(i)}_{I}$ are $\Lambda^i_{II}$ and $\Lambda^i_{IX}$, and for measurements along the $Y$-axis the nonzero elements of each $\Lambda^{\scriptscriptstyle(i)}_{I}$ are $\Lambda^i_{II}$ and $\Lambda^i_{IY}$.
Therefore, for measurements in the computational ($Z$) basis, non-zero values in $\Lambda^i_{IX}$ and $\Lambda^i_{IY}$ would imply a tilt in the measurement axis away from $Z$.
In our experimental example, the values for the first row in each PTM (rounded to $10^{-4}$) are $$\Lambda^{\scriptscriptstyle(0)}_{I} = (0.5117,  0.0018, -0.0026,  0.4940)$$ and $$\Lambda^{\scriptscriptstyle(1)}_{I} = (0.4883, -0.0018 ,  0.0026, -0.4940).$$
The non-zero values for $\Lambda^i_{IX}$ and $\Lambda^i_{IY}$ suggest that the POVM effect $E_i$ associated with outcome $i$ is not perfectly along $Z$, but also contains some $X$ and $Y$ components.
To see this, we can equate \eq\eqref{eq:born_QI} with the Born rule for a POVM effect $E_i$,
\begin{equation}\label{eq:born_POVM}
    p(i | \rho) = \sum_j \Lambda^{\scriptscriptstyle(i)}_{0j} r_j = \Tr[E_i \rho] ~.
\end{equation}
By expanding $\rho$ in the Pauli basis,
\begin{equation}
    \rho = \frac12 \sum_j r_j \sigma_j ~,
\end{equation}
where $r_0 = 1$ and $\sigma_0 = I$, we see that
\begin{equation}
    \frac12 \Tr[E_i \sigma_j] = \Lambda^{\scriptscriptstyle(i)}_{0j} ~.
\end{equation}
Solving for $E_i$:
\begin{equation}
    E_i = \sum_j \Lambda^{\scriptscriptstyle(i)}_{0j} \sigma_j ~.
\end{equation}
So, in our case
$$E_0 \approx 0.5117I +  0.0018X - 0.0026Y + 0.4940Z$$
and
$$E_1 \approx 0.4883I -  0.0018X + 0.0026Y - 0.4940Z ~.$$

It is worth noting that even though our POVM $\{ E_i \}$ appears to be off-axis, it is small enough that it is probably within our experimental uncertainty, so it may not be physically meaningful. These experiments were performed using dispersive readout on a superconducting qubit \cite{siddiqi2006dispersive}, such that the frequency shift of the resonator depends on the qubit state in the $Z$ basis; so, the measurement axis is inherently along $Z$. However, we note that off-axis measurements have been observed in superconducting qubits \cite{chen2019detector}, so it cannot be entirely ruled out in the current data.

\subsection{Post-Measurement State}

The other half of the QI that we must consider is the post-measurement quantum state.
Recalling that the post-measurement state $\rho_i$ conditioned on the classical outcome $i$ [\eq\eqref{eq:post_meas_state}] is
$$\rho_i = \frac{\E_i(\rho)}{\Tr[\E_i(\rho)]} ~,$$
we may re-write this in the PTM formalism using our notation in the previous section as
\begin{equation}
    \sket{\rho_i} = \frac{ \Lambda^{\scriptscriptstyle(i)} \sket{\rho} }{ \Lambda^{\scriptscriptstyle(i)}_{I} \sket{\rho} } ~.
\end{equation}
To understand what can be interpreted directly from the QI PTMs, suppose we apply our QI to a maximally mixed state, $\sket{\rho} = (1, 0, 0, 0)^\trans$, resulting in the following conditional states (in the ideal case):
\begin{align}
    \sket{\rho_0} &= \frac{ 1 }{ 0.5 } \Lambda^{\scriptscriptstyle(0)} \begin{pmatrix}
        1 \\ 0 \\ 0 \\ 0
    \end{pmatrix} 
    = \begin{pmatrix}
        1 \\ 0 \\ 0 \\ 1
    \end{pmatrix} \\ \nonumber
    \\
    \sket{\rho_1} &= \frac{ 1 }{ 0.5 } \Lambda^{\scriptscriptstyle(1)} \begin{pmatrix}
        1 \\ 0 \\ 0 \\ 0
    \end{pmatrix} 
    = \begin{pmatrix}
        1 \\ 0 \\ 0 \\ -1
    \end{pmatrix} ~.
\end{align}
They are exactly the $\ket{0}$ and $\ket{1}$ states, respectively. 
Thus, the left column of $\Lambda^{\scriptscriptstyle(i)}$ tells us exactly what the post-measurement state is, when the input state carries no quantum information.
In this case, the measurement outcome itself completely determines the output state.
For our noisy measurement, we instead observe that 
\begin{align}
    \sket{\rho_0} &\approx \frac{ 1 }{ 0.51 } \begin{pmatrix}
        0.51 \\ 0 \\ -0.01 \\ 0.49
    \end{pmatrix} = \begin{pmatrix}
        1 \\ 0 \\ 0.02 \\ 0.96
    \end{pmatrix} ~, \label{eq:m&p0} \\ \nonumber
    \\ 
    \sket{\rho_1} &\approx \frac{ 1 }{ 0.49 } \begin{pmatrix}
        0.49 \\ 0 \\ 0 \\ -0.41
    \end{pmatrix} = \begin{pmatrix}
        1 \\ 0 \\ 0 \\ -0.84
    \end{pmatrix} ~. \label{eq:m&p1}
\end{align}
These correspond to a vector that is approximately $\ket{0}$ when the outcome is $0$, and a vector that points mostly toward $\ket{1}$ when the outcome is $1$, but is clearly not a pure state; this is consistent with $T_1$ relaxation during measurement, which shrinks the Bloch sphere toward $\ket{0}$. 
Thus, similar to how we can observe the effects of non-unital errors (e.g., $T_1$ decay) in the first column of PTMs introduced in the previous section, non-unital errors also show up in the first column of the PTMs for QIs, although there may appear asymmetries depending on the outcome $i$.

When the input to a QI is the maximally-mixed state, it makes any QI look like a \emph{measure-and-prepare} process \cite{Horodecki2003} --- that is, a process in which the output state $\ket{i}$ is entirely determined by the measurement outcome $i$.
However, measure-and-prepare processes (which are rank-1 matrices; see the ideal PTMs in \fig\ref{fig:qi}) cannot fully model MCMs (which can be higher rank matrices; see the experimental PTMs in \fig\ref{fig:qi}), because in a generalized setting the post-measurement state can retain some coherence beyond what the classical outcome $i$ tells you.
Indeed, in an ideal world, the input state before a MCM should have some coherence (it should not be a mixed state), and the values in the unital block of the QI PTMs (the lower-right $3\times3$ matrix) captures how coherences are mapped from input state to output state.
To see this, consider the post-measurement state when we measure 0 (1) when the input state is $\ket{0}$ ($\ket{1}$):
\begin{align}
    \sket{\rho_0} &= \Lambda^{\scriptscriptstyle(0)} \begin{pmatrix}
        1 \\ 0 \\ 0 \\ 1
    \end{pmatrix} 
    \approx \begin{pmatrix}
        1 \\ 0 \\ -0.02 \\ 0.98 
    \end{pmatrix} ~, \\ \nonumber
    \\
    \sket{\rho_1} &= \frac{ 1 }{ 0.98 } \Lambda^{\scriptscriptstyle(1)} \begin{pmatrix}
        1 \\ 0 \\ 0 \\ -1
    \end{pmatrix} 
    \approx \begin{pmatrix}
        1 \\ 0 \\ 0 \\ -0.85
    \end{pmatrix} ~.
\end{align}
The post-measurement state in each case is approximately $\ket{0}$ and $\ket{1}$, respectively, but not exactly.
Indeed, comparing these vectors to those we would obtain for an ideal QI, where $\sket{\rho_0} = (1, 0, 0, 1)^\trans$ and $\sket{\rho_1} = (1, 0, 0, -1)^\trans$, we can see that our QI is approximately a measure-and-prepare process, but that the post-measurement state is not entirely determined by the measurement outcome $i$ (up to any uncertainties in our experimental outcomes).

This analysis reveals that much can be gleaned from visual inspection of the unital blocks of the QI PTMs.
Firstly, as discussed in the previous section, the diagonal elements of $\Lambda$ quantify how polarization is preserved along each Pauli axis\footnote{This analysis is somewhat more complex when the error channel is not purely Pauli (i.e., for PTMs that are not diagonal), since coherence errors can preserve total polarization, but might rotate elements from one axis to another. Moreover, it is also more complex for QIs, where the PTMs are separated based on the measurement outcome.}.
For example, comparing the diagonal elements ($\Lambda^{\scriptscriptstyle(i)}_{PP}$) of the ideal and experimental PTMs, we observe that coherences along $X$ and $Y$ ($\Lambda^{\scriptscriptstyle(i)}_{XX}$ and $\Lambda^{\scriptscriptstyle(i)}_{YY}$, respectively) are almost entirely destroyed by the QI (they are at most $\sim 0.01$); this is indicative of a purely dephasing channel, which makes sense because a $Z$-basis measurement should fully dephase the transverse coherences of an input state.
Moreover, the absence of any significant off-diagonal elements in the unital block (which appear at most at the level of $\sim 0.01$) suggests that any coherent errors during measurement are small (but still non-zero); these residual coherent errors are likely the reason that the post-measurement state does not fully align with $\ket{0}$ or $\ket{1}$.
Finally, we observe that the $ZZ$ element is nearly ideal for the measure 0 process ($\Lambda^{\scriptscriptstyle(0)}_{ZZ} = 0.49$), but is less than ideal for the measure 1 process ($\Lambda^{\scriptscriptstyle(1)}_{ZZ} = 0.43$); this is further evidence for $T_1$ decay during measurement, which will not preserve $Z$ polarization when the input state is $\ket{1}$.
Thus, from just looking at the experimental PTMs, we can say that the QI is nearly an ideal dephasing channel, but also contains small coherent rotations and significant non-unital errors from $T_1$ decay.

\section{Conclusions} \label{sec:conclusions}

Understanding QIs is necessary to understand what limits the performance of quantum circuits containing MCMs. 
While proxy measures such as classical assignment (readout) fidelity provide some quantitative guidance on the performance of MCMs, they do not capture the coherences preserved or destroyed by an MCM, or the repeatability of such measurements.
For example, a MCM should be quantum non-demolition (QND) \cite{Braginsky1980, Lupacu2007, Braginsky1996quantum} --- that is, repeated measurements should produce identical outcomes, and should not change the expectation value of the measured observable. 
This is extremely important for applications like quantum error correction (QEC), where QND measurements are important for repeated parity checks \cite{stricker2022characterizing}.
While the \emph{QNDness} of a measurement can be characterized using repeated measurements \cite{hazra2025benchmarking} or detector tomography \cite{Pereira2022Complete, pereira2023parallel}, on its own it is not particularly useful for non-rank-1 measurements because it captures repeatability but is not sensitive to whether a measurement impacts other observables. 
For this reason, a full tomographic reconstruction of QIs is the only way to truly understand MCMs.

In this Note, we provide a brief introduction to QIs and discuss how errors in MCMs show up in the PTMs associated with the different measurement outcomes of a QI. 
Interpreting PTMs in general can be an opaque process, and the joint quantum-classical nature of QIs makes this interpretation more difficult.
However, as we discussed in the previous section, much can be understood by direct visual inspection of the different blocks of the PTMs of a QI.
This analysis can be taken a step further by considering the \emph{error generators} of a QI \cite{wysocki2026detailed}, enabling direct quantification of error rates from different physical sources in a QI.
Such information is invaluable as we look for ways to improve the performance of MCMs.

While tomography of QIs can be performed on systems consisting of more than just a single qubit \cite{govia2023randomized}, QI tomography does not scale to systems of arbitrary size (for the same reasons that quantum process tomography does not scale). For this reason, many are considering randomized benchmarking approaches for characterizing MCMs \cite{govia2023randomized, zhang2025generalized, hines2025pauli, hothem2025measuring}. Similar to randomized benchmarks for quantum gates \cite{PRXQuantum.6.030202}, these approaches are much more scalable than QI tomography, but obscure many of the underlying error sources and generally only report high-level metrics such as fidelities, measurement-induced error rates, and error rates of Pauli observables.
Perhaps, given the underlying physics of MCMs, such high-level metrics will be sufficient for characterizing MCMs at scale. 
Indeed, these approaches are only truly descriptive of uniform stochastic instruments \cite{mclaren2023stochastic}---QIs with error models consisting only of stochastic errors. Measurements where errors are already predominantly stochastic, such as dispersive measurement on superconducting qubits, might be amenable to such reduced error models.
However, much remains to be understood regarding the dynamics of MCMs and their impact on nearby spectator qubits, so only time will tell whether randomized benchmarks of MCMs are predictive of actual circuit performance\footnote{They generally are not for quantum gates, so it would be reasonable to expect they are not for MCMs either}.
\bigskip
\paragraph*{\textbf{\textup{Acknowledgements}}}\label{sec:acknowledgements}
A.H.~acknowledges fruitful discussions with Piper Wysocki and Bram Evert.

\clearpage

\bibliography{bibliography}

\end{document}